\documentclass[superscriptaddress,reprint,amsmath,amssymb,aps]{revtex4-1}
\usepackage[english]{babel}
\usepackage[utf8]{inputenc}
\usepackage{amsmath}
\usepackage{amssymb}
\usepackage{mathrsfs}
\usepackage{braket}
\usepackage{physics}
\usepackage{graphicx}
\usepackage{gensymb}

\usepackage[colorinlistoftodos]{todonotes}
\usepackage{siunitx} 
\usepackage{dsfont}
\usepackage[caption=false]{sub fig}

\usepackage{amsmath}
\usepackage[linktocpage,colorlinks=true,linkcolor=blue,citecolor=blue,breaklinks=true]{hyperref}
\usepackage{verbatim}
\usepackage{breakcites}

\usepackage{ulem}

\renewcommand{\vec}[1]{\boldsymbol{#1}}

\newcommand{\bnabla}{\vec{\nabla}}

\newcommand{\beginsupplement}{%
        \setcounter{table}{0}
        \renewcommand{\thetable}{S\arabic{table}}%
        \setcounter{figure}{0}
        \renewcommand{\thefigure}{S\arabic{figure}}%
     }

\begin{document}

\title{Reconfigurable Flows and Defect Landscape of Confined Active Nematics}
\author{Jérôme Hardoüin}
\affiliation{Departament de Quimica Fisica; Universtitat de Barcelona, 08028 Barcelona Spain}
\affiliation{Institute of Nanoscience and Nanotechnology; Universtitat de Barcelona, 08028 Barcelona Spain}
\author{Rian Hughes}
\affiliation{The Rudolf Peierls Centre for Theoretical Physics, Clarendon Laboratory, Parks Road, Oxford, OX1 3PU, United Kingdom}
\author{Amin Doostmohammadi}
\affiliation{The Rudolf Peierls Centre for Theoretical Physics, Clarendon Laboratory, Parks Road, Oxford, OX1 3PU, United Kingdom}
\author{Justine Laurent}
\affiliation{Laboratoire Gulliver, UMR CNRS 7083, ESPCI Paris, PSL Research University}
\author{Teresa Lopez-Leon}
\affiliation{Laboratoire Gulliver, UMR CNRS 7083, ESPCI Paris, PSL Research University}
\author{Julia M Yeomans}
\affiliation{The Rudolf Peierls Centre for Theoretical Physics, Clarendon Laboratory, Parks Road, Oxford, OX1 3PU, United Kingdom}
\author{Jordi Ignés-Mullol}
\affiliation{Departament de Quimica Fisica; Universtitat de Barcelona, 08028 Barcelona Spain}
\affiliation{Institute of Nanoscience and Nanotechnology; Universtitat de Barcelona, 08028 Barcelona Spain}
\author{Francesc Sagués}
\affiliation{Departament de Quimica Fisica; Universtitat de Barcelona, 08028 Barcelona Spain}
\affiliation{Institute of Nanoscience and Nanotechnology; Universtitat de Barcelona, 08028 Barcelona Spain}
\date{\today}

\begin{abstract}

Using novel micro-printing techniques, we develop a versatile experimental setup that allows us to study how lateral confinement tames the active flows and defect properties of the microtubule/kinesin active nematic system. We demonstrate that the active length scale that determines the self-organization of this system in unconstrained geometries loses its relevance under strong lateral confinement. Dramatic transitions are observed from chaotic to vortex lattices and defect-free unidirectional flows. Defects, which determine the active flow behavior, are created and annihilated on the channel walls rather than in the bulk, and acquire a strong orientational order in narrow channels. Their nucleation is governed by an instability whose wavelength is effectively screened by the channel width. All these results are recovered in simulations, and the comparison highlights the role of boundary conditions.

\end{abstract}

\maketitle

\section*{Introduction}
Active matter refers to systems composed of self-driven units, such as tissues, bacterial suspensions, or mixtures of biofilaments and motor proteins, that organise their textures and flows autonomously by consuming either stored or ambient free energy ~\cite{Ramaswamy10, Marchetti13}. This distinctive hallmark sometimes conceals another significant, and often unappreciated, feature of active systems: their capability to adapt to the environments where they reside. For example, human cancer cells switch between distinct invasion modes when they encounter constrictions in the crowded environment of stroma~\cite{Friedl12}, and the growth of bacterial biofilms can be directed by their surroundings~\cite{Conrad18}. Moreover, geometrical confinement tends to control active flows, replacing the bulk chaotic flow state often termed active turbulence, by more regular flow configurations. Understanding the subtleties of how this occurs will have relevance to possible future applications of active materials in microfluidics and self assembly, and in assessing the relevance of the concepts of active matter in the description of biological systems.

Recent contributions dealing with confinement in bacterial suspensions and cell layers have demonstrated a rich range of behaviour. Competition between wall orientation, hydrodynamic interactions, topology and activity lead to a wide variety of flow patterns: spiral vortices~\cite{Wioland13,Lushi14}, synchronised vortex lattices~\cite{Wioland16}, unidirectional flows~\cite{Wioland16NJP,Deforet14,Xi17}, shear flows~\cite{Silberzan18} and freezing~\cite{Duclos14,Duclos17}. Work on confining active mixtures of microtubules and motor proteins to circular domains~\cite{Guillamat17, Guillamat17a, Guillamat17c}, in vesicles~\cite{Keber14} or droplets~\cite{Suzuki17PNAS, Guillamat18} and to more complex geometries such as tori~\cite{Ellis17,Wu17}, have already probed the specific effects of interfacial viscosity, curvature and 3D confinement. These experimental results have prompted parallel simulations~\cite{Miha2013, Whitfield14,Giomi14, Sknepnek15, Zhang16, Alaimo17}.

Here we concentrate on active nematics confined to two-dimensional channels. The system is composed of self-propelling elongated units formed by bundled microtubules powered by ATP-consuming kinesin~\cite{Sanchez12}. Early theoretical work predicted that laterally-confined active nematics undergo an instability to spontaneous laminar flow when the channel width reaches a typical length scale that depends on the strength of the activity~\cite{Voituriez05}. This prediction has been recently confirmed in experiments with spindle-shaped cells~\cite{Silberzan18}. On the other hand, simulations have predicted that, at higher activities or in wider channels, a structured `dancing' state can be stable in active nematics~\cite{Shendruk17}. Our aim here is to assess, in a well-controlled and tunable experimental system, and with the support of numerical simulations, the role of confinement in the patterns and dynamics of an active nematic. In particular, we explore the emergence of a new length scale different from the active length that characterizes the unconfined systems.

\begin{figure*}

\includegraphics[width=\textwidth]{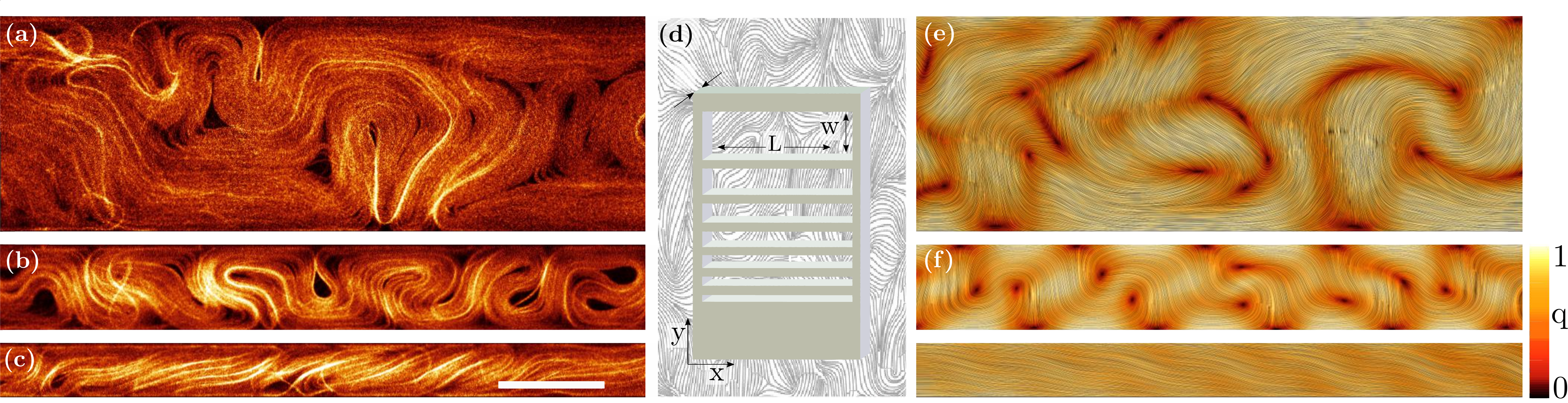}
\caption{
\textbf{Flow States.} (a-c) Confocal fluorescence micrographs of an active nematic interface confined in channels of different widths.  \textit{Scale bar:~$100 \mathrm{\mu m}$}. (d) Top view of the experimental setup including the relevant spatial dimensions. A polymer plate with rectangular openings is placed, by means of a micropositioner, at the interface between the active fluid and silicon oil, thus constraining the existing active nematic. (e-g) Corresponding simulations of the experimental system. Streaking patterns follow the director field tangents. They are produced using a Line Integral Convolution of the director field \cite{loring2014screen} with Paraview software.The color map corresponds to the computed nematic order parameter, $q$. Panels a and e correspond to the active turbulence regime. Panels b and f illustrate the effects of moderate confinement, forcing a new dynamical regime of the defects. Panels c and g correspond to strong confinement, where the filaments are organized into a unstable alignment regime.}
\label{fig:phase_diagram}
\end{figure*}

We find a rich dynamical behaviour, summarised in Fig.~\ref{fig:phase_diagram} and witness the emergence of regimes that have not been previously reported, nor predicted by theories or simulations of active liquid crystals in confinement. More specifically, we uncover a defect-free regime of shear flow in narrow channels. This regime is unstable with respect to the nucleation of short-lived defects at the walls. By increasing the channel width, defect lifetime increases, developing a spatio-temporal organization that corresponds to the predicted state of dancing vortical flows ~\cite{Shendruk17}, before full disorganization into the active turbulence regime for still wider channels, as is typical of the unconfined active nematic. We stress the close interplay between the velocity field and the defect dynamics, and highlight the emergence of a new length scale that, contrary to the classical active length scale, does not depend on the activity level but merely on geometrical parameters.

The paper is organised as follows: we first present the experimental system and the corresponding computational setup. We then describe the different dynamical flow states, and the interplay between flow and the formation of topological defects.  A summary concludes the paper.

\section*{Experimental and computational setup}

The active system we use comprises microtubules powered by ATP-consuming, two headed kinesin molecular motors~\cite{Sanchez12}.
Addition of the depleting agent polyethylene glycol (PEG) concentrates the microtubules into bundles, hundreds of microns long. Within each bundle the kinesin  motors bridge neighbouring microtubules and walk towards their positive ends leading to internal sliding of filaments of opposite polarity. The active nematic was prepared using an open-cell design ~\cite{Guillamat16PRE}, in which
$2 \mathrm{\mu L}$ of the active aqueous microtubule-mixture was placed inside a custom-made pool of $5\mathrm{mm}$ diameter and was covered with $60 \mathrm{\mu L}$ of $100 \mathrm{cS}$ silicon oil (see Materials and Methods). Within $30$ minutes, an active nematic layer extends over the whole surface of the pool.
Driven by the motors, the microtubule bundles at the interface continuously extend and buckle. This gives rise to a dynamical steady state, termed active turbulence, characterised by high vorticity and by the creation, translation and destruction of topological defects.

\begin{figure*}[b]
\includegraphics[width=\textwidth]{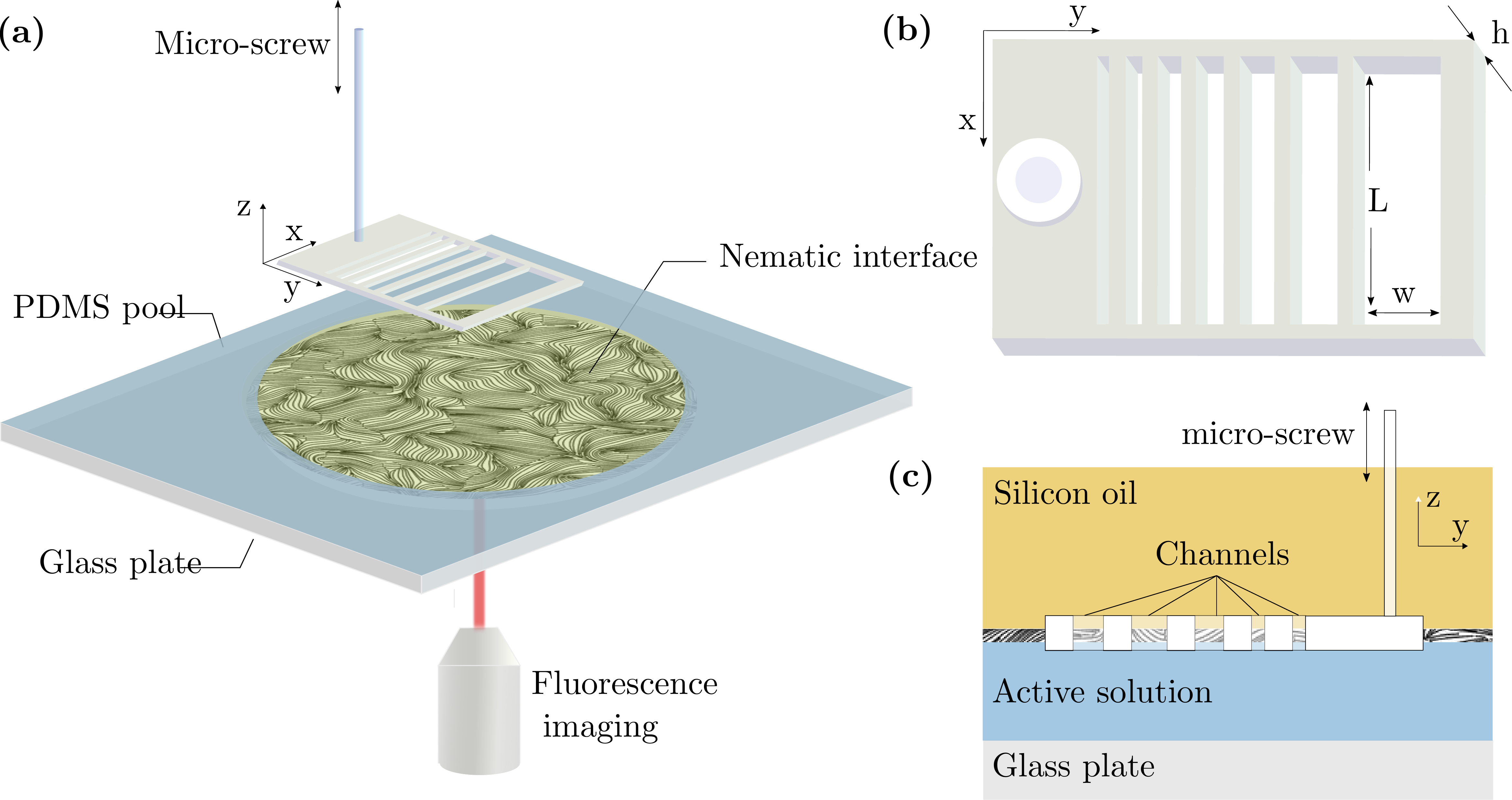}
\caption{\textbf{Experimental setup}. \textbf{(a)} A polymer grid with rectangular openings is placed, by means of a micropositioner, in the custom-made elastomer pool that contains the aqueous active fluid and the passive oil.   \textbf{(b)} Top view with a sketch of the grid including the relevant spatial dimensions. \textbf{(c)} The grid is placed in contact with the oil/water interface. In contact with the grid, the active nematic layer forms only inside the grid openings.}
\label{fig:setup}
\end{figure*}

The layer of active nematic is confined in rectangular enclosures by means of micro-printed polymer grids of $100\mu$m thickness that are placed in contact with the oil/aqueous interface (see Fig. \ref{fig:phase_diagram}). The active nematic is in contact with the active bulk solution underneath, thus ensuring activity and material parameters are equal in all channels. A detailed sketch of the experimental protocol is available in Fig. \ref{fig:setup}.

\section*{Computational Setup}

To model the dynamics of the confined microtubule-motor system we use a continuum description of a two-dimensional, active gel~\cite{Ramaswamy10, Marchetti13,Prost15,Doost18}. The fields that describe the system are the total density $\rho$, the velocity $\vec{u}$, and the nematic tensor $\mathbf{Q} = 2q (\mathbf{nn} - \mathbf{I}/2)$, that describes both the orientation ($\mathbf{n}$) and the magnitude ($q$) of alignment of the nematogens.


The nematic tensor is evolved according to the Beris-Edwards equation~\cite{BerisBook}
\begin{align}
\left(\partial_t + \vec{u}\cdot\bnabla\right) \mathbf{Q} - \mathbf{S} &= \Gamma_{Q} \mathbf{H},
\label{eqn:lc}
\end{align}
where $\mathbf{S}=\xi \mathbf{E}-(\mathbf{\Omega}\cdot\mathbf{Q}-\mathbf{Q}\cdot\mathbf{\Omega})$
is a generalised advection term, characterising the response of the nematic tensor to velocity gradients. Here, $\mathbf{E}=(\bnabla\vec{u}+\bnabla\vec{u}^{T})/2$ is the strain rate tensor, $\mathbf{\Omega} = (\bnabla\vec{u}^{T}-\bnabla\vec{u})/2$ the vorticity tensor, and $\xi$ is the alignment parameter representing the collective response of the microtubules to velocity gradients.
$\Gamma_{Q}$ is a rotational diffusivity
and the molecular field $\mathbf{H} = -\frac{\delta \mathcal{F}}{\delta \mathbf{Q}} + \frac{\mathbf{I}}{2} {\rm Tr} \left(\frac{\delta \mathcal{F}}{\delta \mathbf{Q}}\right)$, models the relaxation of the orientational order to minimise a free energy $\mathcal{F}$.

The free energy includes two terms.  The first is an elastic free energy density,  $\frac{1}{2}K(\bnabla\mathbf{Q})^{2}$, which penalises any deformations in the orientation field of the nematogens and where we assume a single elastic constant $K$.  We note that the free energy functional does not include any Landau-de Gennes bulk free energy terms: all the ordering in the simulations arises from the activity~\cite{Thampi15}. This is motivated by the fact that there is no equilibrium nematic order in the experimental system without ATP (i.e.\  in the absence of active driving). The second contribution to the free energy is a surface anchoring,  $\frac{1}{2}\mathcal{W}\text{Tr}(\mathbf{Q}-\mathbf{Q}_{D})^2$. To correspond to the experiments $\mathbf{Q}_{D}$ is chosen so that the director prefers to align parallel to the boundary walls. The strength of anchoring at the boundaries, $\mathcal{W}$, is set to values corresponding to  weak anchoring so that the nematogens can re-orientate at the walls to allow defects to form there.

The total density $\rho$ satisfies the continuity equation and the velocity $\bf {u}$ evolves according to 
 \begin{align}
 \rho (\partial_t + \vec{u}\cdot\bnabla)\vec{u} &= \bnabla\cdot\mathbf{\Pi},
\label{eqn:ns}
 \end{align}
where $\boldsymbol{\Pi}$ is the stress tensor.
The stress contributions comprise the active stress  $\mathbf{\Pi}^{\text{active}} =  -\zeta\mathbf{Q} $ where $\zeta$ is the activity coefficient, viscous stress $\mathbf{\Pi}^{\text{viscous}} = 2 \eta\mathbf{E}$, where $\eta$ is the viscosity,
and the elastic stresses $\mathbf{\Pi}^{\text{elastic}}=-P\mathbf{I} -2\xi q_{}\mathbf{H}+ \mathbf{Q}\cdot\mathbf{H} - \mathbf{H}\cdot\mathbf{Q}-\bnabla \mathbf{Q} \frac{\delta \mathcal{F}}{\delta \bnabla \mathbf{Q}}$,
where $P=p-\frac{K}{2}(\bnabla\mathbf{Q})^2$ is the modified pressure.
Eqs.~(\ref{eqn:lc})--(\ref{eqn:ns}) were solved numerically using a hybrid lattice-Boltzmann method~\cite{Davide07,Thampi14c}. In the experiments the microtubules slide over the walls and therefore free-slip boundary condition were imposed on the velocity field.
See Materials and Methods for simulation parameters.

\begin{figure*}
\includegraphics[width=\textwidth]{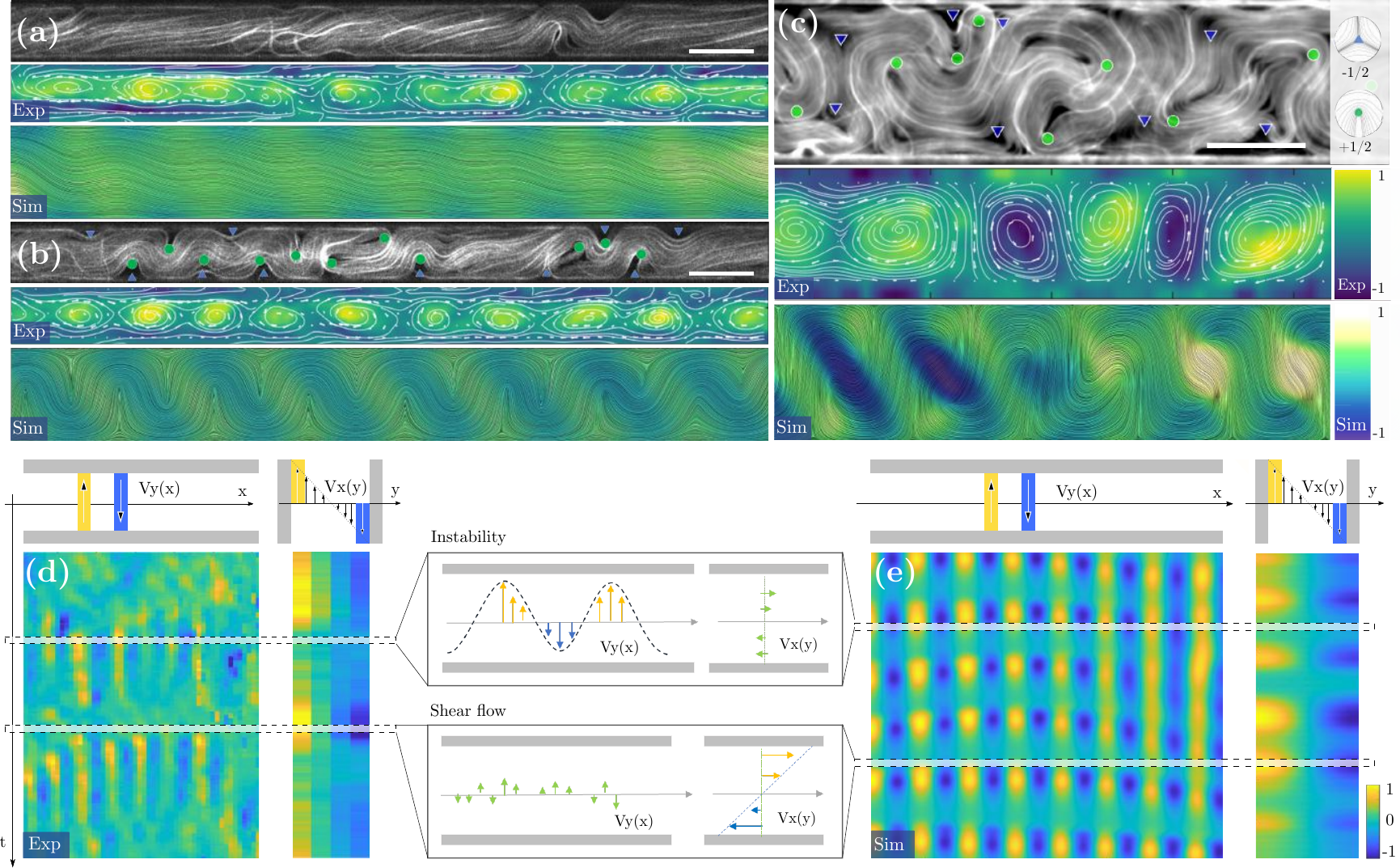}
\caption{(a-d) \textbf{Flow patterns}. 
Each panel is composed of (top) a snapshot of a typical confocal fluorescence image of the active nematic, with defect locations overlaid, (center) experimental PIV measurement of the flow patterns, colored by the normalized vorticity, with velocity streamlines overlaid and (bottom) numerical simulations colored by the normalized vorticity. Lines correspond to the director field. (a) Shear flow state, $\lambda/w\sim 1$, in the shear phase. (b) Shear flow state, $\lambda/w\sim 1$, in the instability phase. (c) Dancing state, $\lambda/w \sim 0.7$. \textit{Scale bar: $50~\mathrm{\mu m}$}. (d-e) \textbf{Periodic instabilities that disrupt the shear state.} Experimental kymographs for $V_x$ averaged along the channel and $V_y$ averaged across the channel are shown in (d) while the corresponding simulations appear in (e). The color encodes the normalized velocity components. One instance of stable shear flow regime and one instance of transversal instability are sketched.
\label{fig:patterns}
}
\end{figure*}

\begin{figure}[h]
\begin{center}
\includegraphics[width=\columnwidth]{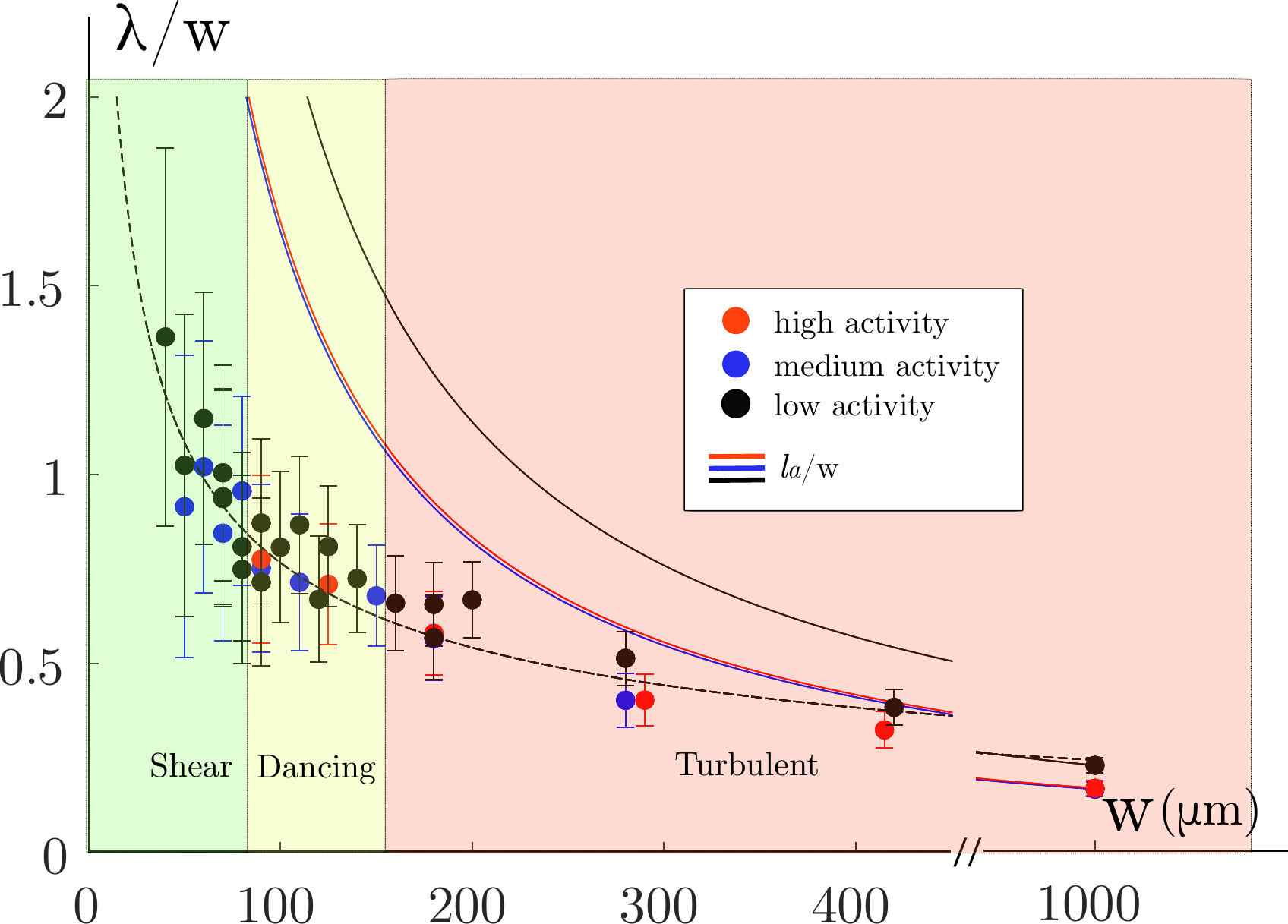}
\caption{\textbf{Defect Spacing.} Experimental mean defect spacing, $\lambda$, rescaled by the channel width, $w$, as a function of $w$. Different colors correspond to different activities. The error bars correspond to the standard deviation for measurements at 10 random times for each experiment. The continuous lines correspond to $\lambda/w=l_a/w$ for each experiment, with $l_a$ being the active length scale corresponding to the value of $\lambda$ in the unconfined case.
\label{fig:SI_spacing}}
\end{center}
\end{figure}

\begin{figure*}
\begin{center}
\includegraphics[width=\textwidth]{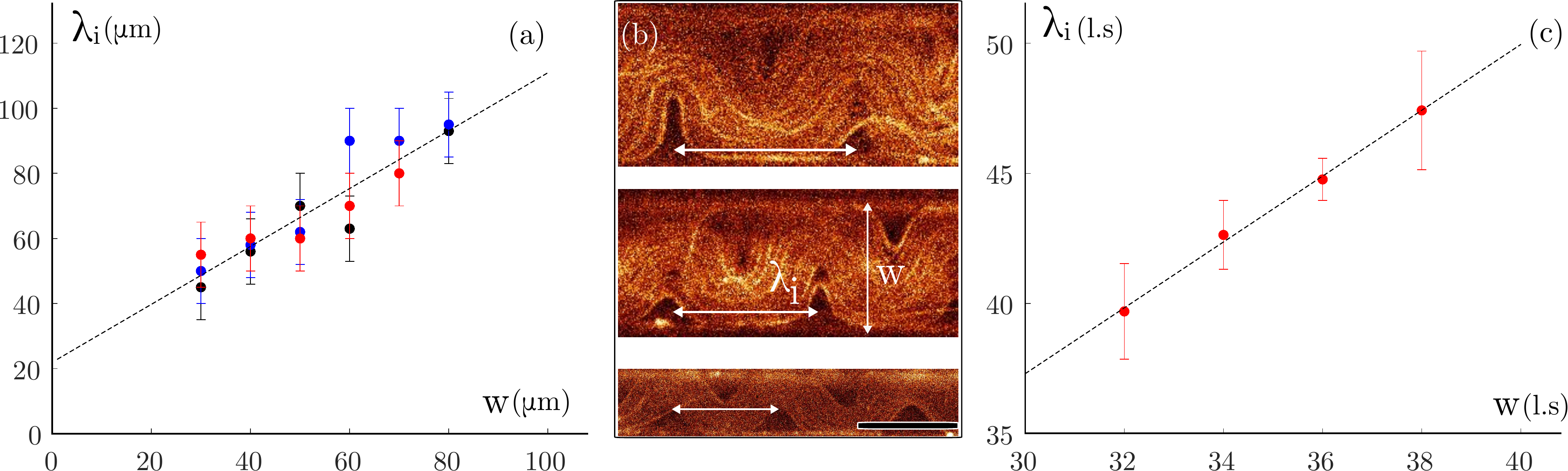}
\caption{\textbf{Instability wavelength}.(a) Measurement of the instability wavelength $\lambda_i$ {\it vs} $w$ for different values of activity, controlled by the ATP concentration, $c_0=1.55~\mathrm{mM}$ (red), $c_0/5$ (blue) and $c_0/10$ (black). (b) Measurement of the instability wavelength $\lambda_i$ at the onset of defect nucleation in the shear state for three different channels widths $w$ and for an activity $c_0$. $\lambda_i$ is taken as the distance between two neighbouring negative defects along a given wall, as indicated. \textit{Scale bar:~$50 \mathrm{\mu m}$}. (c) $\lambda_i$ {\it vs} $w$ in simulations, for a single value of activity. The unit of the axes is the number of lattice sites (l.s). In (a), error bars correspond to an estimated measurement uncertainty of $10~\mathrm{\mu m}$ for all data points. In (c) error bars correspond to the standard deviation of the measurements.
\label{fig:SI_wavelength}}
\end{center}
\end{figure*}

\section*{Results and Discussion}

The main experimental control parameters of our system are the channel width, $w$, and the concentration of ATP, which determines the activity.  In this section, we describe experiments and simulations showing how both the defect landscape and flow patterns evolve as $w$ is increased. We identify two well defined regimes: a shear flow regime, observed for $w <80~\mathrm{\mu m}$, which is transiently defect-free (Fig. \ref{fig:patterns} (a-b) and \textbf{Videos S1-2}), and the dancing defects regime, for $w >90~\mathrm{\mu m}$ (Fig. \ref{fig:patterns} (c) and \textbf{Videos S3A-B}). The transition between these two regimes is not sharp. For values of $w$ in the range $80-90~\mathrm{\mu m}$, the direction of the shear is not uniform along the channel, but rather composed of patchy domains where shear flow spontaneously arises with a random direction, and is then quickly disrupted by instabilities (see \textbf{Videos S4A-B}).



We find that the relevant parameter setting the dynamic state in the system is $\lambda/w$, where $\lambda$ is the mean defect separation. The latter is defined as $\lambda=({Lw}/{N})^{1/2}$, where $L$ is the length of a given channel, $w$ its width, and $N$ the number of defects averaged in time. For unconfined active nematics, this length scale coincides with $l_a=\sqrt{K/\zeta}$, often referred to as the active length-scale, which determines the vortex size distribution and corresponds to the mean defect separation~\cite{Thampi14c,Giomi15,Guillamat17}. However, we find that, in confined active nematics, $\lambda$ is no longer equal to $l_a$. Instead, it significantly decreases with the channel width (see Fig. \ref{fig:SI_spacing}).

%

%

\subsection*{A defect-free state: shear flow disrupted by instabilities}

\begin{figure*}
\begin{center}
\includegraphics[width=0.5\textwidth]{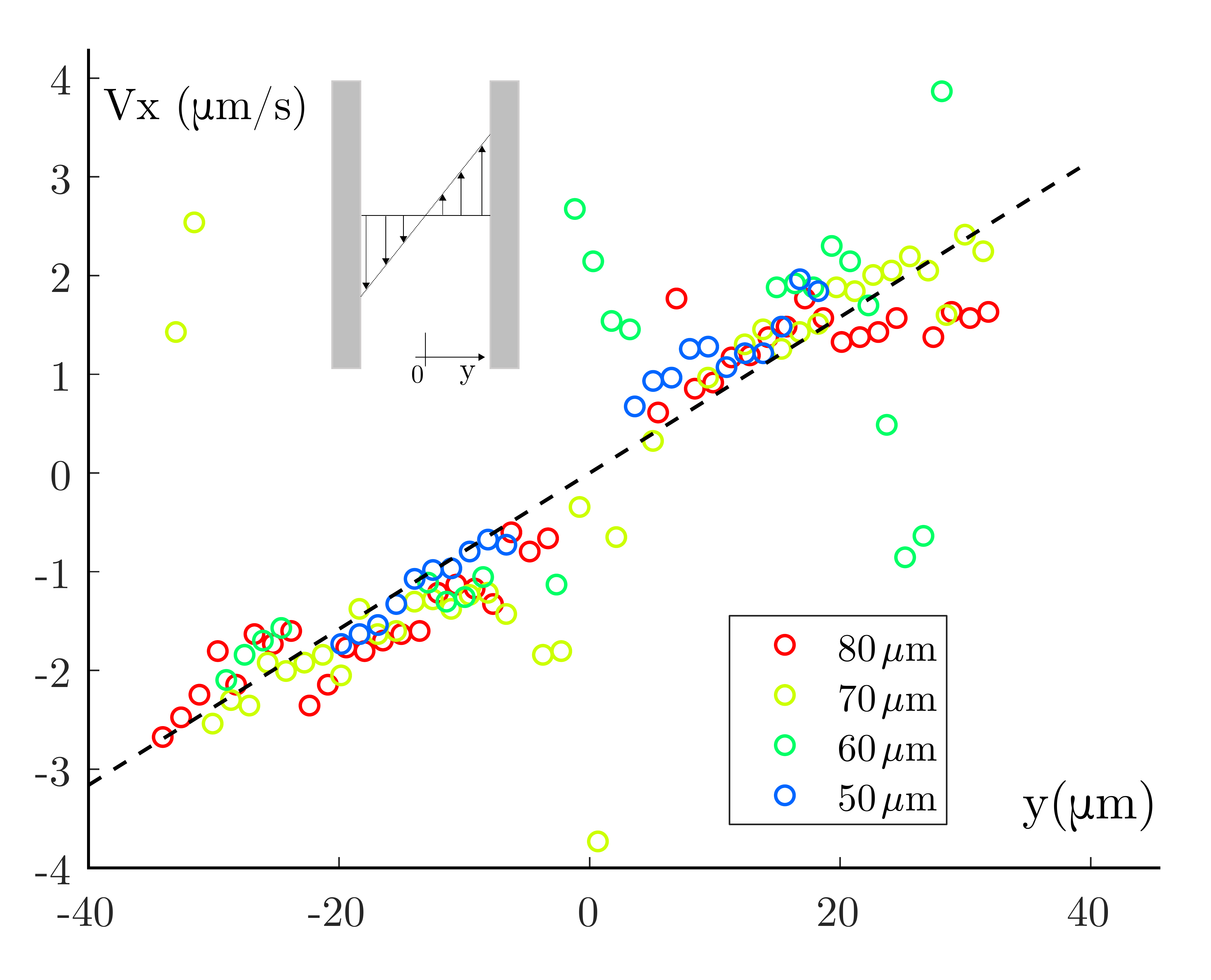}
\caption{\textbf{Shear flow profile.} Experimental flow profile of $\mathrm{V_x}$ across the channel. Different colors refer to different channel widths. All channels are plotted on the same graph with no rescaling, the origin of the y-axis is taken at the center of each channel. Velocities are computed with a custom ImageJ plugin (see materials and methods for details).
\label{fig:SI_profile}}
\end{center}
\end{figure*}

The \textit{shear flow regime} is a defect free state that appears for $\lambda/w > 1$ (Fig. \ref{fig:SI_spacing}), which is experimentally realized for $w < 80~\mathrm{\mu m}$. The active material is primarily aligned parallel to the walls over distances that can persist along the whole channel as shown in Fig. \ref{fig:patterns} (a). A global shear deformation is observed, with flows along the channel (Fig. \ref{fig:SI_profile}). The maximum velocities are measured at the walls, with opposite signs, and the velocity perpendicular to the walls is negligible. The shear rate, characterised by the slope of the velocity profile, is approximately constant over a relatively large range of channel widths. This is as expected because the activity, and hence the energy input per unit area, which must be balanced by the viscous dissipation due to the shear, is the same for all channels.

Extensile, aligned active nematics are intrinsically unstable to bend deformations \cite{Simha02, Voituriez05, Rao07, Guillamat16}. As a consequence, the sheared state eventually leads to local bend instabilities as shown in Fig. \ref{fig:patterns} (b). As a result, the velocity field repeatedly switches between two different states: longitudinal shear flow and a transversal instability regime (see {\bf Videos S1A-B}). The instability takes the form of a sinusoidal deformation of the aligned nematic field, with a well defined length-scale along the channel. As the perturbation progresses, defects are rapidly nucleated from the walls, at regularly-spaced positions  coinciding with the maxima of the sinusoidal perturbation.  We measured the wavelength of the instability and found that it scales with the channel width (Fig. \ref{fig:SI_wavelength}). This is strong evidence that the hydrodynamics is screened, and that the channel width is important in controlling the flows.

\begin{figure}
\includegraphics[width=\columnwidth]{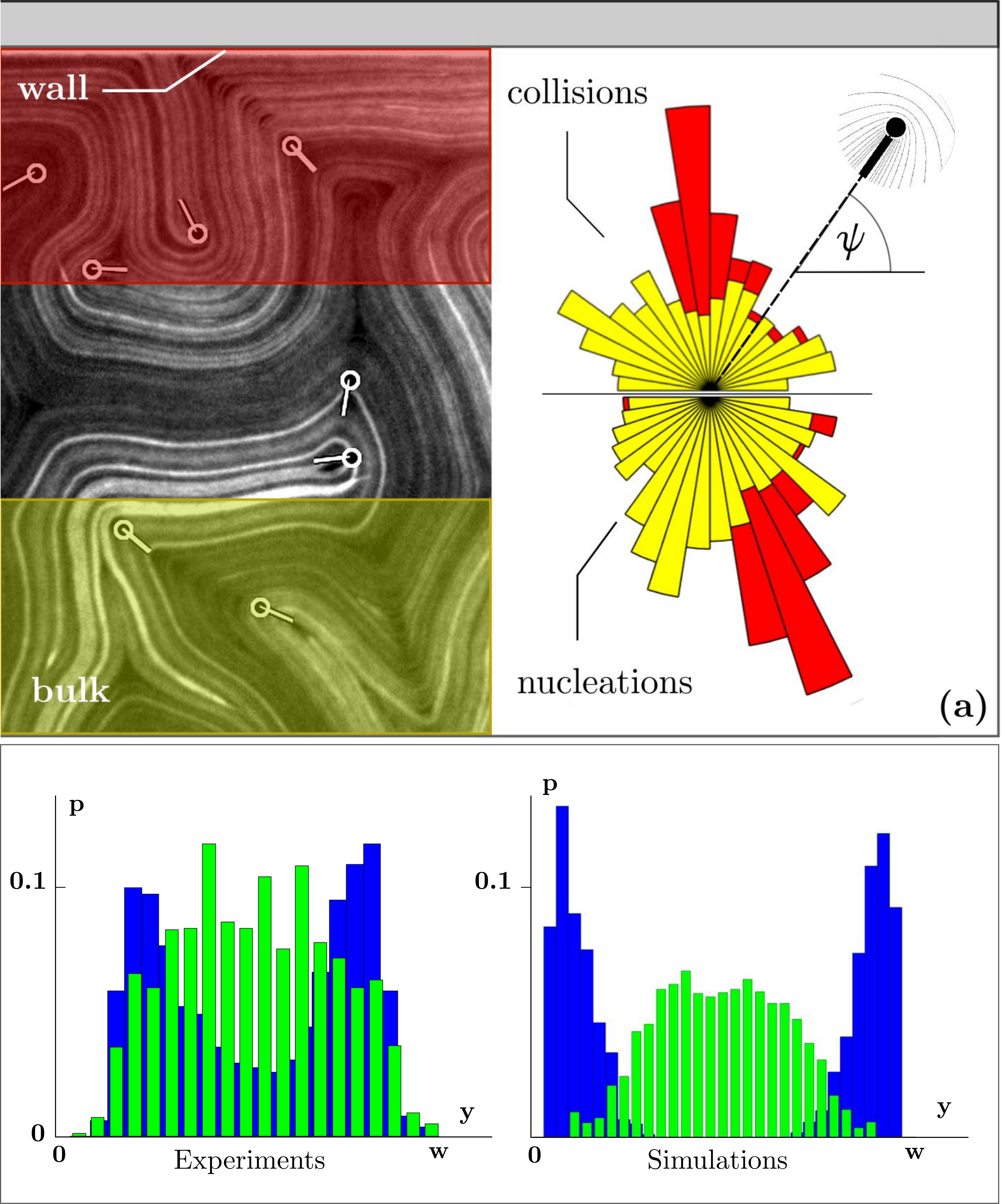}
\caption{\textbf{Defect nucleation.} (a) Statistical distribution of positive defect orientations as a function of the distance from the wall. The angle $\psi$ corresponds to the orientation of the defect with respect to the wall. $\psi=\pi/2$ (resp. $\psi=3\pi/2$) corresponds to a defect perpendicularly colliding with (resp. moving away from) the wall. (b) Experimental (left) and simulated (right) probability distributions of defect position across a channel. Green (blue) distribution refers to positive (negative) defects.
\label{fig:wall_effects}
}

\end{figure}

Upon their nucleation at the boundaries, the orientation of the $+1/2$ defects is strongly anisotropic (see Fig. \ref{fig:wall_effects}(a)). They preferentially align perpendicular to the walls and, due to their active self-propulsion~\cite{Sanchez12,Pismen13,Giomi14}, they move away from the walls into the bulk. On the contrary, because of their three-fold symmetric configuration, $-1/2$ defects have no self-propulsion and remain in the vicinity of the walls \cite{Shendruk17}.
Eventually, the $+1/2$ defects reach the opposite wall and annihilate with negative defects residing close to it. In this way, the defect-free phase is periodically restored (see \textbf{Video S2}). Remarkably, even though no chirality is observed in the sheared state, as we repeat the experiments, the handedness of the shear flow initially selected is preserved through successive instability cycles. This memory can be explained by observing that the instability is triggered locally and that it is entrained by the neighboring sheared regions.

The dynamics of the switching behavior and the coexistence of the shear and the instability states can be best illustrated in a space-time diagram of the averaged velocity components in the channel, as shown in Fig. \ref{fig:patterns}  (d-e). In the shear state, the velocity perpendicular to the boundaries and averaged across the channel width, $\langle V_y(x)\rangle_y$, vanishes, while the velocity component parallel to the channel walls and averaged along the channel length, $\langle V_x(y)\rangle_x$, is maximum at the walls. Once bend instabilities are triggered, defect pairs form and $+1/2$ defects propagate across the channel and dismantle the shear state, which leads to the emergence of non-zero values of $\langle V_y(x)\rangle_y$ with a well-defined length scale along the channel. The defects eventually reach the opposing wall and annihilate, such that the shear state is reestablished. As is apparent from Fig. \ref{fig:patterns}  (d-e), over time the active system alternates between the two regimes.
Simulations allowed to test channel widths well below the experimental capabilities, allowing to explore the $\lambda \gg w $ regime.  For these conditions, we observed stable shear flow, as any pairs of defects that were generated at the channel walls immediately self-annihilated (see \textbf{Video S5}). \\
\subsection*{The dancing state: a one-dimensional line of flow vortices}

\begin{figure*}
\begin{center}
\includegraphics[width=0.7\textwidth]{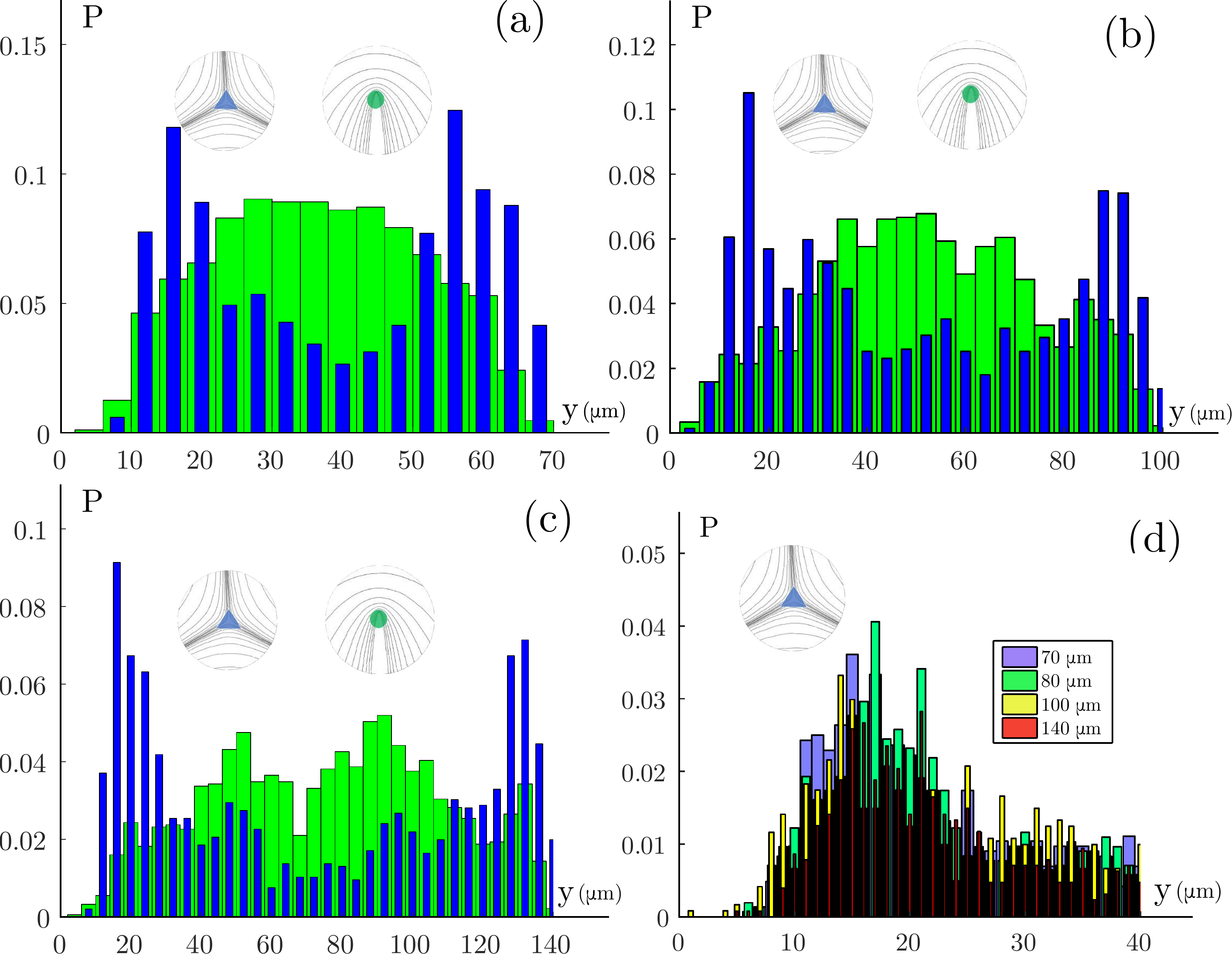}
\caption{\textbf{Defect Unbinding.} Experimental probability distribution of topological defect positions across the channel width. (a), (b) and (c) correspond to the profiles for different channel widths. Green (resp blue) curves stand for +1/2 (resp -1/2) defects. (d) Distribution of $-1/2$ defects position close to a wall. The position of the wall is given by $y=0$. Different colors stand for different channel widths. The maxima of the distributions all coincide at a distance around $18~\mathrm{\mu m}$ from the wall.
\label{fig:SI_distribution_exp}}
\end{center}
\end{figure*}

Increasing the channel width to values between 90$\mu m$ and 120$\mu m$, one-dimensional arrays of vortices are observed as shown in Fig. \ref{fig:patterns} (c). A close look at the defect distribution reveals that the transition to the flow state with organised vorticity arrays corresponds to the point when the channel can accommodate more than one defect in its cross-section, {\it i.e.}, $\lambda/w < 1$ (Fig. \ref{fig:SI_spacing}). One could expect that this criterion is reached when the channel width becomes comparable to the active length-scale. However as shown in Fig. \ref{fig:SI_spacing}, $l_a/w$ is still much larger than $1$ in the range of the dancing state, \textit{i.e.}, $\lambda\ll l_a$. Futhermore, contrary to the $l_a/w$ curves, $\lambda/w$ does not seem to depend on activity, supporting the idea that $\lambda$ is indeed a pure geometrical feature. Dynamically, the system behaves as if two distinct populations of positive defects are travelling along the channel in opposite directions, passing around each other in a sinusoidal-like motion. A similar state had been predicted by simulations and was referred to as the {dancing state} \cite{Shendruk17}. However, contrary to the published simulations, the dancing state observed in our experiments is quite fragile, and vortex lattices are always transient and localised in space.  Defects may annihilate with their negative counterparts, or even switch their direction of motion, thus perturbing the flow pattern (see \textbf{Videos S3A-B}).
The difference between the spatial organisation of oppositely charged defects in the confined active nematic is manifest in their arrangement across the width of the channel (Fig. \ref{fig:wall_effects} (b)). The distribution of the $+1/2$ defects has a single peak at the centre of the channel (Fig. \ref{fig:wall_effects} (b), green). On the other hand, the $-1/2$ defect distribution has two peaks, one at each of the boundary walls (Fig. \ref{fig:wall_effects}(b), blue) and the profiles do not rescale with the channel width. Instead, the wall-peak distance is approximately constant at a separation  $\sim 18 \mathrm{\mu m}$ from the wall, as shown in Fig. \ref{fig:SI_distribution_exp}(d). This can be attributed to $-1/2$ defects having no self-propulsion and thus interacting elastically with the channel walls~\cite{Denniston96}. The distance of the $-1/2$ defect from the walls is therefore expected to be controlled by the intrinsic anchoring penetration length of the nematic $l_{\text{n}}=K/\mathcal{W}$, which is set by the competition between the orientational elastic constant $K$ and the strength of the anchoring at the wall $\mathcal{W}$, and is independent of the channel width and activity of the particles. As expected, for wider channels, $w > 120\mu$m,
the difference between the  +$1/2$ and $-1/2$ defect distributions diminishes as active turbulence is established (see Fig. \ref{fig:SI_distribution_exp}(c)).
We observed a similar behaviour in the simulations, but the -1/2 defects were more strongly localised near the walls, and the +1/2 defects consequently tended to lie towards the centre of the channel (Fig. \ref{fig:wall_effects}(b)).
Simulations have also allowed us to pinpoint the necessary boundary conditions at the channel walls to trigger defect nucleation. We find that such a localised defect formation at the walls is obtained only for free-slip boundary conditions for the velocity, and weak planar anchoring boundary conditions for the nematic director field. This is because the free-slip velocity, together with the parallel anchoring of the director, allows for strong tangential active flows, and hence strong tangential nematic order, to develop along the boundaries.
This results in bend instabilities that grow perpendicular to the walls with the weak strength of the anchoring allowing the director to deviate from a planar configuration at the positions where the bend instability is developed.
Previous simulations that assumed no-slip velocity and strong alignment conditions on the confinement did not observe defect nucleation at the boundaries~\cite{Shendruk17,Norton17}, and  showed insensitivity of the active nematic patterns to the boundary conditions~\cite{Norton17}. This is because the strong anchoring used in these works prevented defects forming at the walls.

It is, however, interesting to note that a recent computational study, based on a kinetic approach, has reported a special case of defect nucleation at the boundaries of an active nematic confined within a circular geometry with no-slip velocity boundary condition and free anchoring~\cite{Gao17}. However, here the wall-bound defect nucleation was observed only for confining disks of small sizes and had a very different dynamics than that reported here: the +1 defect imposed by the circular geometry was first dissociated into two $+1/2$ defects and then, for sufficiently small confinement one of the $+1/2$ defects kept moving into and out of the boundary. This is in contrast to our  results  that show regularly-spaced defect pair nucleation sites at the boundaries for a range of  channel widths. Moreover, the corresponding defect spacing is governed by an instability wavelength that is no longer given by the conventional active length scale.


\section*{SUMMARY}

We have presented experimental results, supported by continuum simulations, investigating the flow and defect configurations of an active nematic confined to rectangular channels of varying width. Our experiments have identified a new dynamical state, where well-defined shear flow alternates in a regular way with bursts of instability characterised by $+1/2$ topological defects moving across the channel. We have also shown that, for wider channels, it is possible to identify the dancing state \cite{Shendruk17}, although the particular boundary conditions considered in the present work make it less stable.

Our work highlights the importance of topological defects in controlling the confined flows. Because the microtubules have weak planar anchoring and can freely slide along the channel walls, pairs of $\pm 1/2$ defects form at the walls of the channel. The $+1/2$ defects are self-propelled and move away from the walls whereas the $-1/2$ defects remain close to the boundaries. The distance to the boundaries is set by the anchoring penetration length. In bulk active nematics the defect spacing is set by the active length scale and, although there is some evidence of long-range ordering \cite{DeCamp15, Vromans16, Srivastava16}, defect motion is primarily chaotic. In confinement, however, the defect spacing and the wavelength of the instability are set by the channel width and the defect trajectories are more structured.

Together, experiments and simulations demonstrate a surprisingly rich topological defect dynamics in active nematics under channel confinement, and a sensitive dependence on both channel width and boundary conditions. Therefore, confinement provides a way of controlling active turbulence and defect trajectories, a pre-requisite for using active systems in microfluidic devices.

\section*{Materials and Methods}
Complementary information about the preparation of active samples and grid manufacturing is given in Supporting Information, together with additional comments on image acquisition and data analysis. Details pertaining to the simulations including a table with all parameter values and a discussion on boundary conditions are also provided.

\section*{Acknowledgments}
We are indebted to the Brandeis University Materials Research Science
and Engineering Centers (MRSEC) Biosynthesis facility for providing the tubulin. We thank
M. Pons, A. LeRoux, and G. Iruela (University of Barcelona) for assistance in the expression of
motor proteins. J.H., J.I.-M and F.S. acknowledge funding from Ministerio de Economia, Industria y
Competitividad, Spain (project FIS2016-78507-C2-1-P, Agencia Estatal de Investigaci\'on/European Regional Development Fund)
J.H. acknowledges funding from the European Union's
Horizon 2020 research and innovation program under grant agreement no. 674979-
NANOTRANS.
S. Brandeis University MRSEC Biosynthesis facility is supported by
NSF MRSEC DMR-1420382. T.L-L. acknowledges funding from the French Agence Nationale de la Recherche (Ref. ANR-13-JS08-006-01). Gulliver institute acknowledges funding from $2015$ Grant SESAME MILAMIFAB (Ref. 15013105) for the acquisiton of a Nanoscribe GT Photonic Professional device.
R.H. acknowledges that this research was supported/partially supported by the European Union’s Horizon 2020 research and innovation programme under the Marie Sklodowska-Curie grant agreement No 722497 - LubISS.
\section*{Author Contributions}
J.H, J.L, F.S, and J.I.-M. conceived the experiments and R. H., A.D, and J.M.Y. designed the simulations. J.H. performed the experiments in collaboration with T.L.-L., and analyzed the experimental data. R.H conducted the numerical simulations. J.H, R.H and J.M.Y. wrote the manuscript with contribution from all the authors.
\section*{Competing Interests}
The authors declare that they have no competing interests.

\section*{Materials and Methods}

\subsection*{\textbf{Protein preparation}}
Microtubules (MTs) were polymerized from heterodimeric ($\zeta,\beta$)-tubulin
from bovine brain [a gift from Z. Dogic’s group at Brandeis University (Waltham, MA)], incubated at $37 \degree \mathrm{C}$ for 30 min in aqueous M2B buffer (80 mM Pipes, 1 mM EGTA, 2 mM MgCl2) prepared with Milli-Q water. The
mixture was supplemented with the reducing agent dithiothrethiol (DTT)
(Sigma; 43815) and with guanosine-5-[($\zeta,\beta$)-methyleno]triphosphate (GMPCPP)
(Jena Biosciences; NU-405), a slowly hydrolysable analog of the biological
nucleotide guanosine-5\'-triphosphate (GTP) that completely suppresses the
dynamic instability of the polymerized tubulin. GMPCPP enhances
spontaneous nucleation of MTs, obtaining high-density suspensions of
short MTs ($1$–$2~\mathrm{\mu m}$). For fluorescence microscopy, 3\% of the tubulin was
labeled with Alexa 647. Drosophila melanogaster heavy-chain kinesin-1
K401-BCCP-6His (truncated at residue 401, fused to biotin carboxyl carrier
protein (BCCP), and labeled with six histidine tags) was expressed in
Escherichia coli using the plasmid WC2 from the Gelles Laboratory (Brandeis
University) and purified with a nickel column. After dialysis against
500 mM imidazole aqueous buffer, kinesin concentration was estimated by
means of absorption spectroscopy. The protein was stored in a $60\%$ (wt/vol)
aqueous sucrose solution at $-80 \degree \mathrm{C}$ for future use.
\subsection*{\textbf{Imaging}}
Images were acquired using a laser scanning confocal microscope Leica TCS SP2 AOBS with a $10\times$ objective at typical frame rate of 1 image per second. For each experiment and at each time, a frame is acquired in both fluorescence and reflection mode. Fluorescence is used to visualize the nematic field, and reflection images are used for Particle Image Velocimetry (PIV) measurements as explained in the data analysis section.
\subsection*{\textbf{Grid Manufacturing}}
The grids are printed using a two-photon polymerization printer, a Nanoscribe GT Photonic Professional device, with a negative-tone photoresist IP-S (Nanoscribe GmbH, Germany) and a $25\times$ objective. The grids were directly printed on silicon substrates without any preparation to avoid adhesion of the resist to the substrate (plasma cleaner of the substrate, for example, would increase the adhesion).
After developping 30 minutes in Propylene Glycol Monomethyl Ether Acetate (PGMEA 99,5\%, Sigma Aldrich) and 5 minutes in isopropanol (Technical, VWR), a batch polymerization is performed with UV-exposure (5 min at 80\% of light power). After printing onto a silicon wafer, the grids are bound to a vertical glass capillary with a UV-curable glue. The capillary is then delicately manipulated to detach the grids from the printing support. The grids are washed in three steps (iso-propanol, DI water, ethanol) and dried with a nitrogen stream before each experiment. The thickness of the grids is $100 \mathrm{\mu m}$, to ensure good mechanical resistance. We have used grids with rectangular openings 1.5 mm long and widths ranging from 30 to 300$\mu$m. Each grid contains different channel widths to ensure that simultaneous experiments can be performed with the same active nematic preparation, thus ensuring that material parameters remain unchanged when comparing different confinement conditions.

\subsection*{\textbf{Active Gel preparation}}
Biotinylated kinesin motor protein and
tetrameric streptavidin (Invitrogen; 43-4301) aqueous suspensions were incubated
on ice for 30 min at the specific stoichiometric ratio 2:1 to obtain
kinesin–streptavidin motor clusters. MTs were mixed with the motor clusters
that acted as cross-linkers, and with ATP (Sigma; A2383) that drove the activity
of the gel. The aqueous dispersion contained a nonadsorbing polymeric
agent (PEG, 20 kDa; Sigma; 95172) that promoted the formation of
filament bundles through depletion. To maintain a constant concentration
of ATP during the experiments, an enzymatic ATP-regenerator
system was used, consisting on phosphoenolpyruvate (PEP) (Sigma; P7127)
that fueled pyruvate kinase/lactate dehydrogenase (PK/LDH) (Invitrogen;
434301) to convert ADP back into ATP. Several antioxidant components
were also included in the solution to avoid protein denaturation, and to
minimize photobleaching during characterization by means of fluorescence
microscopy. Anti-oxydant solution 1 (AO1) contained $15~\mathrm{mg/mL}$ glucose and $2.5\mathrm{M}$ DTT. Anti-oxidant solution 2 contained $10~\mathrm{mg/mL}$ glucose oxydase (Sigma G2133) and $1.75~\mathrm{mg/mL}$ catalase (Sigma, C40). Trolox (Sigma, 238813) was used as an additional anti-oxidant. A high-salt M2B solution was used to raise the $\mathrm{MgCl_2}$ concentration. The PEG-based triblock copolymer surfactant Pluronic F-127
(Sigma; P-2443) was added to procure a
biocompatible water/oil interface in subsequent steps. Buffer for stock solutions of PEP, DTT, ATP, PEG and Streptavidin was M2B, and we added $20~\mathrm{mM}$ of $\mathrm{K_2HPO_4}$ to the buffer of Catalase, Glucose, Glucose Oxydase and Trolox. A typical recipe is summarized in Table \ref{table:Conc}.

\begin{table}[ht]
\begin{center}
\begin{tabular}{l|l|l}
\textbf{Compound }&\textbf{Stock solution} &$\mathbf{v/V_{total}}$\\ \hline
PEG & $12~\mathrm{\%~w/vol}$& $0.139$   \\
PEP      &$200~\mathrm{mM}$ & $0.139$  \\
High-salt M2B      &$69~\mathrm{mM~MgCl_2}$ & $0.05$  \\
Trolox    &$20~\mathrm{mM}$ & $0.104$     \\
ATP  & $50~\mathrm{mM}$ & $0.03$  \\
Catalase     & $3.5~\mathrm{mg.ml}^{-1}$ &$0.012$\\
Glucose    & $300~\mathrm{mg.ml}^{-1}$ &$0.012$   \\
Glucose Oxydase     & $20~\mathrm{mg.ml}^{-1}$ &$0.012$   \\
PK/LDH   & $600-1000~\mathrm{units.ml}^{-1}$ & $0.03$      \\
DTT     & $0.5~\mathrm{M}$&$0.012$    \\
Streptavidin    & $0.352~\mathrm{mg.ml}^{-1}$ &$0.023$    \\
Kinesin    & $0.07~\mathrm{mg.ml}^{-1}$&$0.234$     \\
Microtubules    & $6~\mathrm{mg.ml}^{-1}$&$0.167$     \\
Pluronic    & $17~\mathrm{\%}$&$0.027$      \\ \hline
\end{tabular}
    \caption{Composition of all stock solutions, and their volume fraction in the final mixture.}
    \label{table:Conc}
    \end{center}
\end{table}


\subsection*{\textbf{Data analysis}}
Statistical information on defect unbinding and defect orientation (Fig. \ref{fig:wall_effects}) were obtained using the defect detection and tracking Matlab codes developed by Ellis et al.~\cite{Ellis17,Wu17}, coupled with a custom-made direction detection Matlab code derived from Vromans and Giomi's method \cite{Vromans16}.
PIV measurements in Fig. \ref{fig:patterns} (a-c) were obtained using confocal images in reflection mode. In this mode, the active nematic layer exhibits textures that can quite efficiently act as tracers for PIV softwares. The images were treated with ImageJ \textit{PIV} plugin \cite{tseng2012spatial}. The data was then processed with custom Matlab codes.
Flow profiles were computed using a custom ImageJ plugin. PIV measurements failed to give reliable measurements close to the walls for narrow channels because of a substantial drop of resolution that we attribute to artefacts in the reflection mode. Because of this, for a system in the \textit{shear flow} state, the longitudinal velocity $V_x(y)$ was determined as follows. For each $y$ position, a kymograph is built by assembling the image profile along the channel ($x$ coordinate) at different times. The inhomogeneous reflected-light intensity results in traces in the $x-t$ plane of the kymograph, whose slope gives $V_x(y)$. The average value of this slope is obtained by computing the FFT of the kymograph image and finding at which angle the FFT has the highest intensity. We repeat the same process for each $y$ coordinate, resulting in the desired flow profile. The kymograph in Fig. \ref{fig:patterns}  (d) was obtained with a custom Matlab code using PIV data from imageJ as an input. Defect spacing in Fig. \ref{fig:SI_spacing} was obtained by manually counting the number of defects in a given channel, for at least 10 random frames. Velocities were computed using imageJ plugin \textit{Manual Tracking}, averaging the velocities of 10 defects per experiment over their lifetime, typically 15 seconds.

\subsection*{\textbf{Simulations}}
The equations (\ref{eqn:lc}-\ref{eqn:ns}) of the manuscript are solved using a hybrid lattice-Boltzmann (LB) method.
A finite difference scheme was used to solve equation (\ref{eqn:ns}) on a five-point stencil
to discretize the derivatives on a square grid. These are coupled to Navier-Stokes equations,
that are solved using lattice-Boltzmann method with the Bhatnagar-Gross-Krook
(BGK) approximation and a single relaxation time for the collision operator \cite{LB_Book}. The
time integration is performed using PECE predictor-corrector method. The discrete space
and time steps are chosen as unity for the LB method. The algorithm is implemented
using the C++ programming language.

All simulations were performed on a $1800\times w$ rectangular grid, where $w$ varied from 30 to 160 lattice sites. The boundary conditions for the velocity field along all walls and corners are free-slip.  The boundary conditions for the director field are weak planar along the channel walls, which is implemented via the free energy term $\frac{1}{2}\mathcal{W}\text{Tr}(\mathbf{Q}-\mathbf{Q}_{D})^2$. The corners have $\mathbf{Q}_{D}$ such that the director field aligns at an angle of 45 degrees to the length of the channel, on the front-right and back-left corners, and at an angle of 135 degrees on the front-left and back-right corners. The left and right walls have no free energy term applied, and have Neumann boundary conditions.

The parameters used in the simulations are given in Table \ref{table:Params}.  Initially, we chose parameters in a range that has previously been successful in reproducing the dynamics of experiments in microtubule bundles \cite{Melaugh,Lloyd}, and then further refined the parameters via a search in phase space. All simulations were run for up to 300,000 simulation time steps. Some simulations were performed up to 1,000,000 time steps to further test the stability of the simulations.

The initial velocity field was set to zero everywhere in the domain. The orientation of the director field was initialised at an angle of either 0, 35 or 90 degrees to the length of the channel, up to some noise. The noise was implemented using the uniform$\_$real$\_$distribution() function in the standard C++ library. In all simulations, the parameters were set to lattice units.

The figures of the simulations in Fig. \ref{fig:phase_diagram} and Fig. \ref{fig:patterns} (a-c) were created using the programme Paraview. In Fig. \ref{fig:phase_diagram}, the director field was overlaid on the order parameter field q. In Fig. \ref{fig:patterns} (a-c), the director field was overlaid on the vorticity field. The kymograph of the velocity components $V_{x}$ and $V_{y}$ in Fig. \ref{fig:patterns}  (d-e) was created using Matlab, by calculating the mean value of the velocity components in a subsection of the channel at any given time. The distributions of the defect positions across the channel width in simulations shown in Fig. \ref{fig:wall_effects} and \ref{fig:SI_distribution_exp} were created using Matlab. Each bar represents the normalised number of defects along a strip of lattice sites parallel to the channel length, of width one lattice site. The supplementary videos of the simulations were created using Paraview, and any defect tracking was created by overlaying the Paraview images with markers from Matlab.

In the simulations, $\lambda_{i}$ was calculated as follows. At any given channel width and time-step, the distance between each +1/2 defect and its nearest neighbour was calculated. The mean of this quantity was then taken, to give the mean defect-defect distance at a fixed time. Then an average over time was taken to give $\lambda_{i}$ for a single simulation. This process was repeated for three different simulations with different initial conditions. The initial conditions were distinguished by the initialisation angle of the director field, which was selected to be 0, 45 and 90 degrees to the length of the channel. The mean value of these $\lambda_{i}$'s was then taken to give the value plotted in Fig. \ref{fig:SI_wavelength}. The standard deviation of these values were used to get the error bars.

\begin{table}[!ht]
\begin{center}
\begin{tabular}{l|l|l}
\textbf{Parameter} & \textbf{Symbol} & \textbf{Value}\\ \hline
Alignment parameter&$\xi$ & 0.9    \\
Anchoring srength&W      & 0.002  \\
Total density&$\rho$    & 40     \\
Rotational diffusivity&$\Gamma_Q$  & 0.4    \\
Elastic constant&K     & 0.015  \\
Viscosity&$\eta$    & 1/6    \\
Channel length&$L_{X}$     & 1800   \\
Channel width&$L_{Y}$     & 30 - 160     \\
Activity&$\zeta$  & 0.0072 - 0.01 \\ \hline
\end{tabular}
    \caption{Parameters used in the simulations. The values for $\zeta$ are 0.0072, 0.0075, 0.0082, 0.0092 and 0.01. All values are in LB units.}
    \label{table:Params}
    \end{center}
\end{table}
%



\clearpage 

\beginsupplement

\section*{Supplementary Material}
\setcounter{page}{1}

\begin{figure*}
\begin{center}
\includegraphics[width=0.7\textwidth]{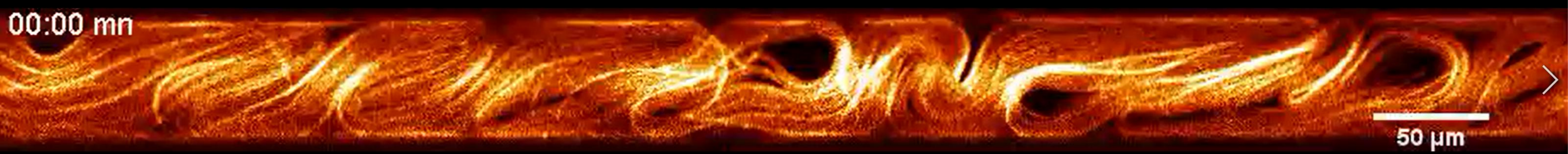}
\caption{\textbf{Video S1A}. Shear flow of active nematics confined in a $50~\mathrm{\mu m}$ channel. The images were acquired with a laser scanning confocal microscope (see materials and methods). \textit{frame rate}: $0.15~\mathrm{fps}$. \textit{scale}:$0.73~\mathrm{\mu m/px}$.
}
\end{center}
\end{figure*}

\begin{figure*}
\begin{center}
\includegraphics[width=0.7\textwidth]{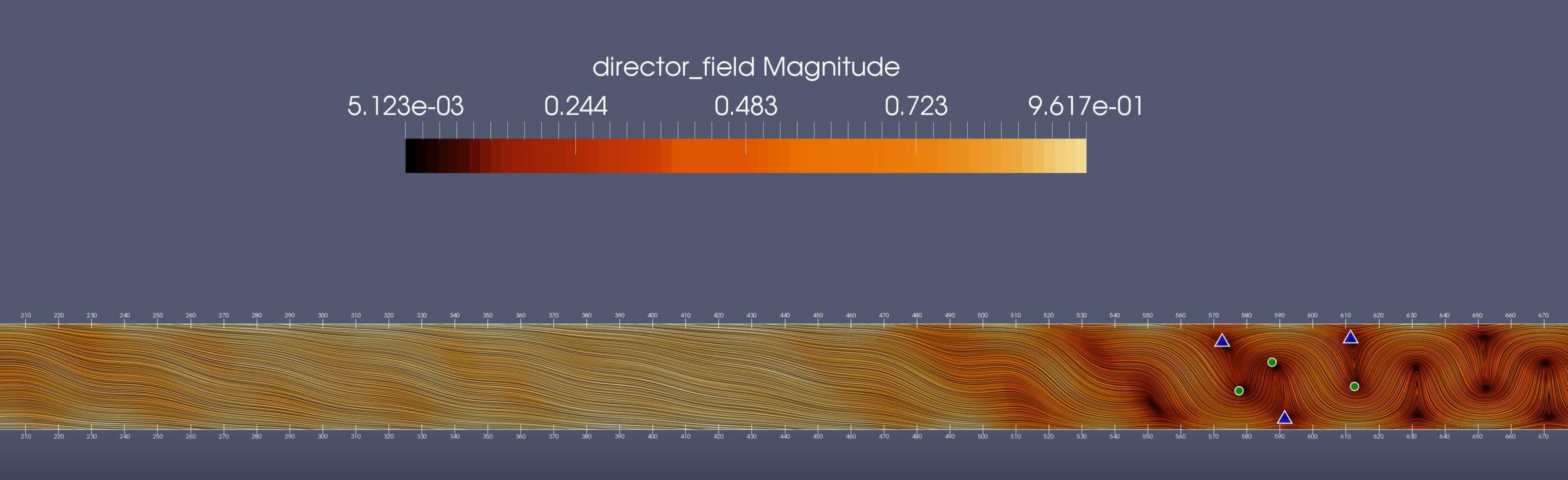}
\caption{\textbf{Video S1B}. Shear state from the simulations. Defects are highlighted by the green circle (+1/2) and the blue triangle (-1/2). Channel width is 32 lattice sites.
}
\end{center}
\end{figure*}

\begin{figure*}
\begin{center}
\includegraphics[width=0.7\textwidth]{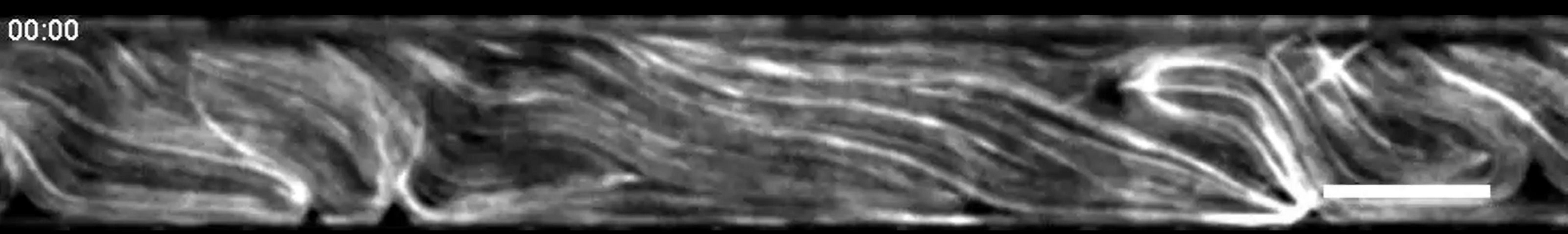}
\caption{\textbf{Video S2}. Shear flow of active nematics confined in a $80~\mathrm{\mu m}$ channel. The images were acquired with a laser scanning confocal microscope (see materials and methods). \textit{frame rate}: $1~\mathrm{fps}$. \textit{scale}:$0.69~\mathrm{\mu m/px}$. \textit{scale bar} $50 \mathrm{\mu m}$.
}
\end{center}
\end{figure*}

\begin{figure*}
\begin{center}
\includegraphics[width=0.7\textwidth]{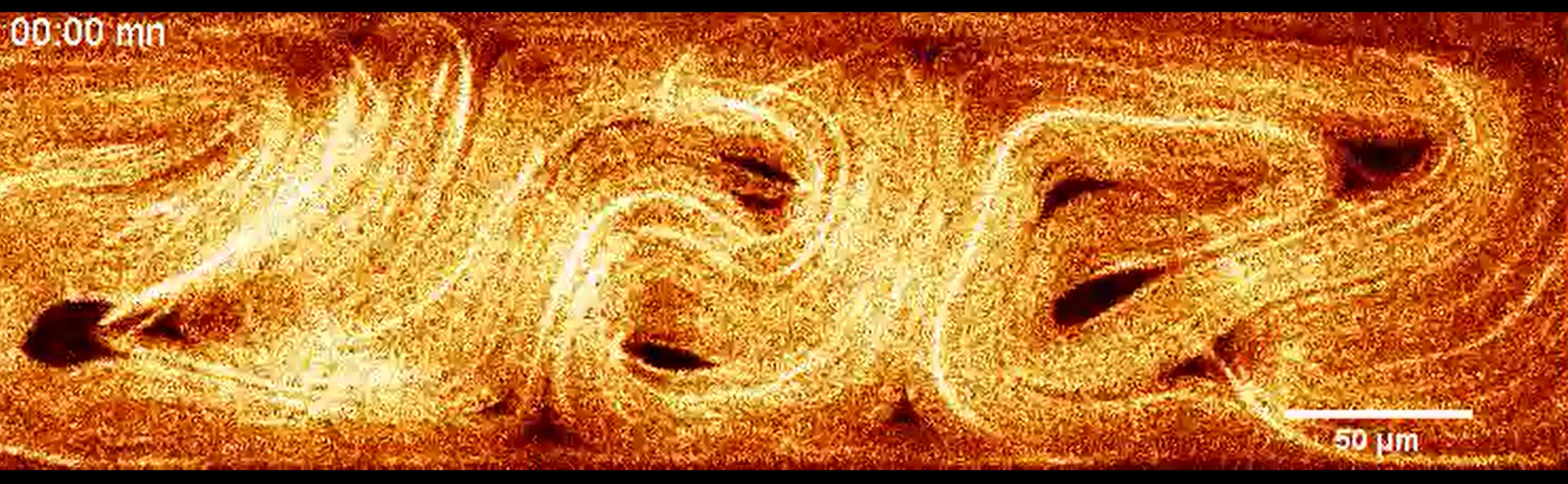}
\caption{\textbf{Video S3A}. Dancing disclination state in a $120~\mathrm{\mu m}$ channel. Positive (+1/2) defects are overlaid with green disks. The images were acquired with a laser scanning confocal microscope (see materials and methods). \textit{frame rate}: $0.6~\mathrm{fps}$. \textit{scale}:$0.56~\mathrm{\mu m/px}$.
}
\end{center}
\end{figure*}
\begin{figure*}
\begin{center}
\includegraphics[width=0.7\textwidth]{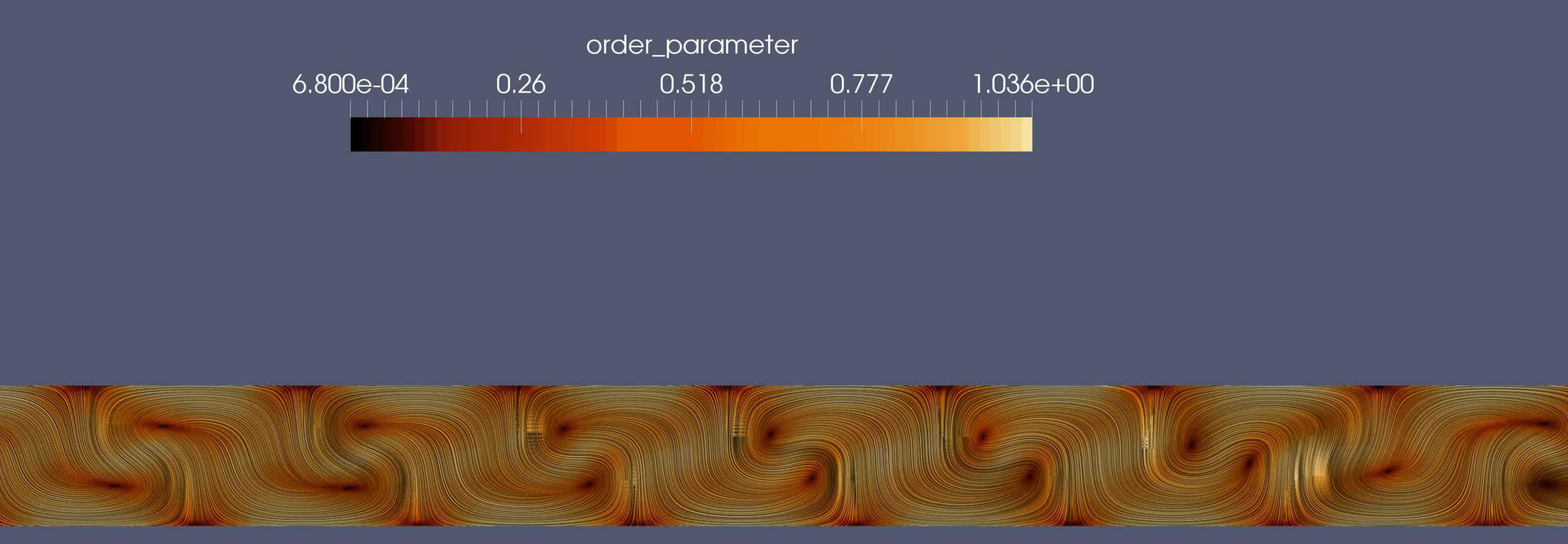}
\caption{\textbf{Video S3B}. Dancing state from the simulations. Channel width is 60 lattice sites.
}
\end{center}
\end{figure*}
\begin{figure*}
\begin{center}
\includegraphics[width=0.7\textwidth]{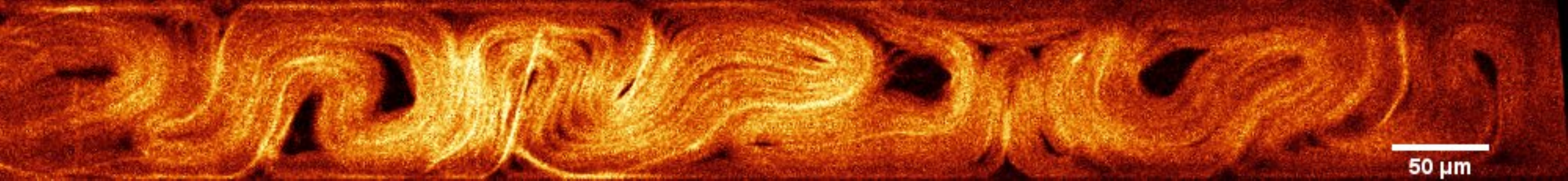}
\caption{\textbf{Video S4A}. Switching shear flow active nematics confined in a $90~\mathrm{\mu m}$ channel. The images were acquired with a laser scanning confocal microscope (see materials and methods). \textit{frame rate}: $0.3~\mathrm{fps}$. \textit{scale}:$0.79~\mathrm{\mu m/px}$.
}
\end{center}
\end{figure*}
\begin{figure*}
\begin{center}
\includegraphics[width=0.7\textwidth]{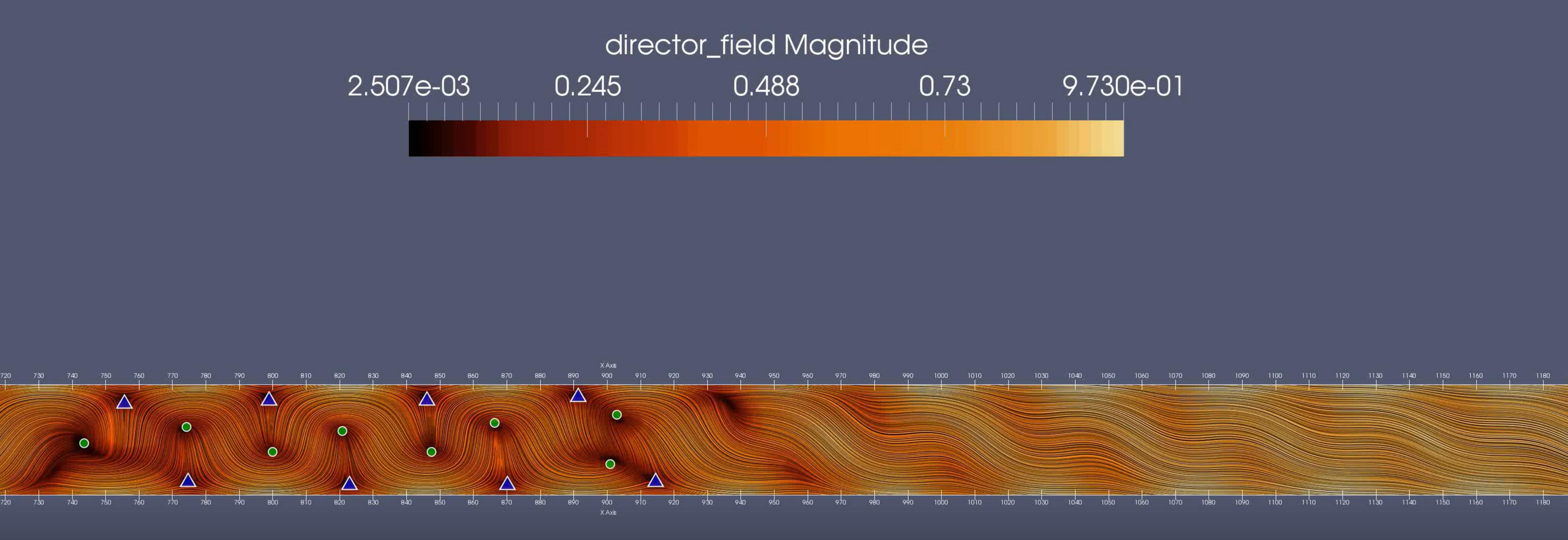}
\caption{\textbf{Video S4B}. Switching state from the simulations. Defects are highlighted by the green circle (+1/2) and the blue triangle (-1/2). Channel width is 40 lattice sites.
}
\end{center}
\end{figure*}
\begin{figure*}
\begin{center}
\includegraphics[width=0.7\textwidth]{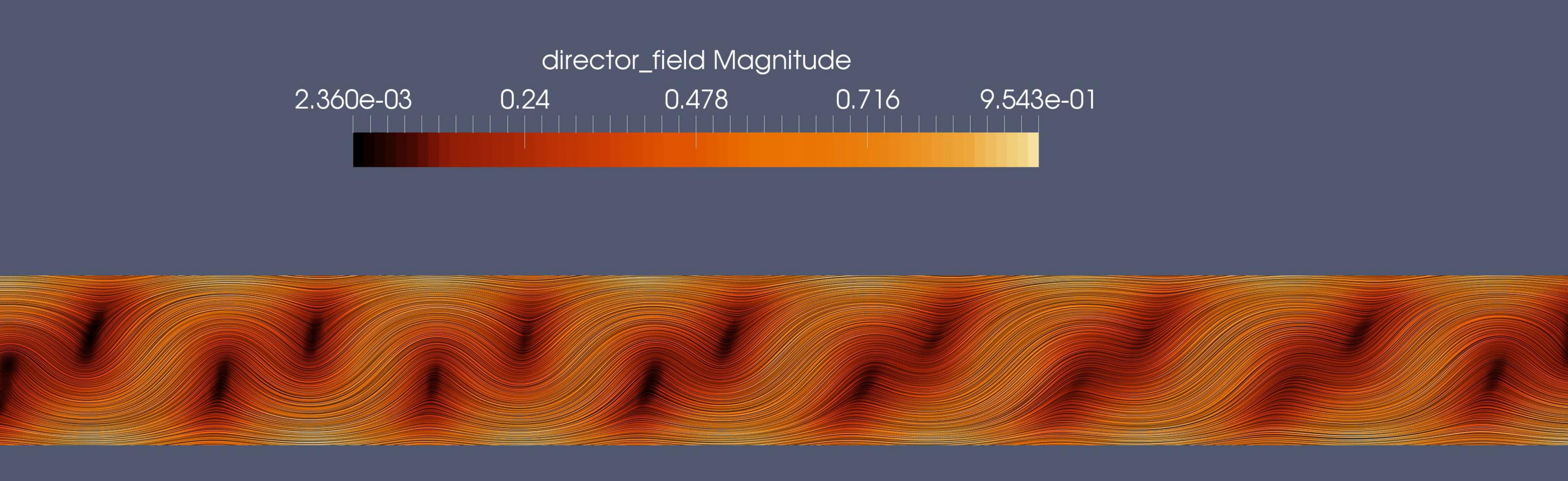}
\caption{\textbf{Video S5}. Self-collapsing state from the simulations. The bend instabilities grow and then collapse in on themselves. Channel width is 30 lattice sites.
}
\end{center}
\end{figure*}

\end{document}